\documentclass[aip, cp, amsmath, amssymb, reprint]{revtex4-2}

\usepackage{graphicx}
\usepackage{dcolumn}
\usepackage{bm}
\usepackage[utf8]{inputenc}
\usepackage[T1]{fontenc}
\usepackage{mathptmx}
\usepackage{xspace}
\usepackage[dvipsnames]{xcolor}

\definecolor{olive}{rgb}{0, 0.7, 0}

\newcommand{\annztwo}{\textsc{annz}$2$\xspace}

\newcommand{\photoz}{photo-$z$\xspace}

\begin{document}

%%%%%%%%%%%%%
%%% TITLE %%%
%%%%%%%%%%%%%
\title{Characterising Improvements in Photometric Redshift Probability Density Functions with Galaxy Morphology}

%%%%%%%%%%%%%%%%%%%%%%%%%%%%%
%%% AUTHOR AND AFFLIATION %%%
%%%%%%%%%%%%%%%%%%%%%%%%%%%%%
\author{John Y. H. Soo}
\email[Corresponding author: ]{johnsooyh@usm.my}
\affiliation{School of Physics, Universiti Sains Malaysia, 11800 USM, Pulau Pinang, Malaysia.}

\author{Benjamin Joachimi}
\email{b.joachimi@ucl.ac.uk}
\affiliation{Department of Physics and Astronomy, University College London, Gower Street, London WC1E 6BT, UK.}

\date{\today}

\begin{abstract}
In this work, we studied the impact of galaxy morphology on photometric redshift (photo-$z$) probability density functions (PDFs). By including galaxy morphological parameters like the radius, axis-ratio, surface brightness and the S\'{e}rsic index in addition to the $ugriz$ broadbands as input parameters, we used the machine learning photo-$z$ algorithm \annztwo to train and test on galaxies from the Canada-France-Hawaii Telescope Stripe-82 (CS82) Survey. Metrics like the continuous ranked probability score (CRPS), probability integral transform (PIT), Bayesian odds parameter, and even the width and height of the PDFs were evaluated, and the results were compared when different number of input parameters were used during the training process. We find improvements in the CRPS and width of the PDFs when galaxy morphology has been added to the training, and the improvement is larger especially when the number of broadband magnitudes are lacking.
\end{abstract}

\maketitle

%%%%%%%%%%%%%%%%%%%%%
%%% INTRODUCTION %%%%
%%%%%%%%%%%%%%%%%%%%%
\section{Introduction} \label{sec:intro}
% Can use \subsection{} and \subsubsection{}.
% Citations: use \cite{} or \onlinecite{}.
% \begin{eqnarray} for multiple numbered equations.
% Use \text{} for Roman text within a math environment.
% Use \begin{subequations} \begin{equation} for equations 1a, 1b etc.

In the recent development of photometric redshifts (\photoz's), the use of probability distribution functions (PDFs, or $p(z)$) has become much sought after. Since the $p(z)$ of a galaxy produced may provide more information than a point estimate \photoz, many have dedicated their time into developing and improving the ways $p(z)$'s are generated \cite{fernandez-soto_error_2002,polsterer_uncertain_2016}. Other than being able to show the probability of multiple peaks, the $p(z)$'s produced for every galaxy in a sample could be stacked together to form a smooth \photoz distribution $n(z)$. Many have found that this form of $n(z)$ has showed better results in weak lensing analyses \cite{gerdes_arborz:_2010,bonnett_using_2015}.

The quality of the $p(z)$'s produced by a \photoz algorithm, however, is difficult to assess: unlike spectroscopic redshifts, there is no 'true' $p(z)$ for a certain galaxy to be compared with. Many recent and ongoing works have been dedicated to introduce and assess quality metrics for $p(z)$'s \cite{wittman_overconfidence_2016,schmidt_evaluation_nodate}. Examples of such quality metrics include the continuous ranked probability score (CRPS), probability integral transform (PIT), and the quantile-quantile (QQ) plot, in which the latter is a graphical representation of the PIT \cite{polsterer_uncertain_2016}.

In this work, we are interested to characterise some of these $p(z)$ metrics by studying if an improvement in the \photoz point estimates would be reflected in an improvement in the $p(z)$ metrics. In particular, we study quantitatively if the improvement in \photoz brought by including galaxy morphological parameters in the training of an artificial neural network (ANN) would reflect an improvement in the $p(z)$ metrics. This is important in its own right, as it helps producers of \photoz's to calibrate their algorithms to produce 'better' $p(z)$'s, so to say.

This work is a quick follow-up study to the work of Soo et al. (2018), in which they showed that galaxy morphology improves the point estimate \photoz's of galaxies in the CS82 sample, and the magnitude of improvement increases with decreasing number of broadband filters used \cite{soo_morpho-z:_2018}. Using this as the basis of our definition of 'improvement', we extend Soo's qualitative analysis on PDFs into a quantitative one.

%%%%%%%%%%%%%%%%%%%%%%%%%%%%%%%%%%%%%%%%
%%% METHODOLOGY AND DATA SAMPLE USED %%%
%%%%%%%%%%%%%%%%%%%%%%%%%%%%%%%%%%%%%%%%
\section{Methodology and Data Sample Used}

In this work, we use the exact same data sample, \photoz algorithm and input parameters used in \cite{soo_morpho-z:_2018}. These would be briefly introduced below, but the reader could refer to \cite{soo_morpho-z:_2018} for more details.

The galaxy sample used is constructed by cross-matching data from various surveys: it uses $ugriz$ broadband photometry from the Sloan Digital Sky Survey (SDSS) Stripe-$82$ Coadd \cite{annis_sloan_2014}, high-quality morphology from the Canada-France-Hawaii Telescope Stripe-$82$ (CS82) Survey \cite{moraes_cfht/megacam_2014}, and spectroscopic redshifts from SDSS, DEEP2, WiggleZ and the VIMOS VLT Deep Survey (VVDS) \cite{york_sloan_2000,newman_deep2_2013,drinkwater_wigglez_2010,le_fevre_vimos_2013}. This sample contains $59498$ galaxies, which is divided equally into $3$ sets for training, validation and testing respectively.

The \photoz algorithm used is \annztwo \cite{sadeh_annz2:_2016}, it is a powerful package capable of utilising several machine learning methods to estimate \photoz point estimates and PDFs of galaxies. Using the same settings as those in \cite{soo_morpho-z:_2018}, several runs of \photoz's for the same set of galaxies are produced, in each run we vary the number of broadband magnitudes used as training inputs (e.g. $i$, $ri$, $gri$, $ugri$, $ugriz$ and etc), in the same fashion shown in Section~$6.1$ of \cite{soo_morpho-z:_2018}. These runs are repeated by adding $5$ morphological parameters as training inputs, these parameters are the galaxy radius ($r$), axial ratio ($q$), mean surface brightness ($\mu$), S\'{e}rsic index ($n$) and shape probability ($p$). These latter runs are known as the 'with morphology' runs, which quality metrics will be compared to the former 'without morphology' runs. In \cite{soo_morpho-z:_2018} it has been established that the \photoz point estimate metrics (root-mean square error, $68$th percentile error and outlier rate) are generally better in the 'with morphology' runs.

We note that the results shown in \cite{soo_morpho-z:_2018} have been reweighted with respect to the CS82 target sample so to reflect its performance on that sample. In this work, however, we consider both the weighted and unweighted cases to see if there are differences in results between the two.

%%%%%%%%%%%%%%%
%%% RESULTS %%%
%%%%%%%%%%%%%%%
\section{Results and Discussion}\label{results}

\subsection{Metrics Used}
To evaluate quantitatively the overall impact of galaxy morphology on the $p(z)$'s produced, we assess the mean value of several metrics and characteristics of each individual galaxy $p(z)$. A total of $5$ metrics will be assessed:

\begin{enumerate}
\item \textit{The mean CRPS value} ($\rho_\textrm{CRPS}$), it tells us how close the position of the true redshift is located to the peak of the $p(z)$, the smaller the value the better the fit \cite{polsterer_uncertain_2016};
\item \textit{The mean PIT value} ($\rho_\textrm{PIT}$), it tells us if the $p(z)$ produced have adequately defined widths, a large value indicates that the $p(z)$'s are either too wide or too narrow \cite{polsterer_uncertain_2016};
\item \textit{The mean Bayesian odds} ($\bar{\Theta}$), a value between $0$ and $1$, it measures the confidence of the PDF produced, the closer to $1$ the more confident and reliable the $p(z)$ \cite{soo_morpho-z:_2018};
\item \textit{The mean PDF height} ($\bar{h}$), which we assume that the higher the better; and
\item \textit{The mean PDF $68$th percentile width} ($\bar{w}_{68}$), which we assume the narrower the better.
\end{enumerate}

We note that these values or metrics are not expected to correlate with one another, e.g. a high $\rho_\textrm{CRPS}$ (accurately peaked) does not imply a low $\rho_\textrm{PIT}$ (correctly shaped).

\subsection{Impact of Galaxy Morphology on $p(z)$}

\begin{figure}
\centering
\includegraphics[width=0.8\linewidth,page=3]{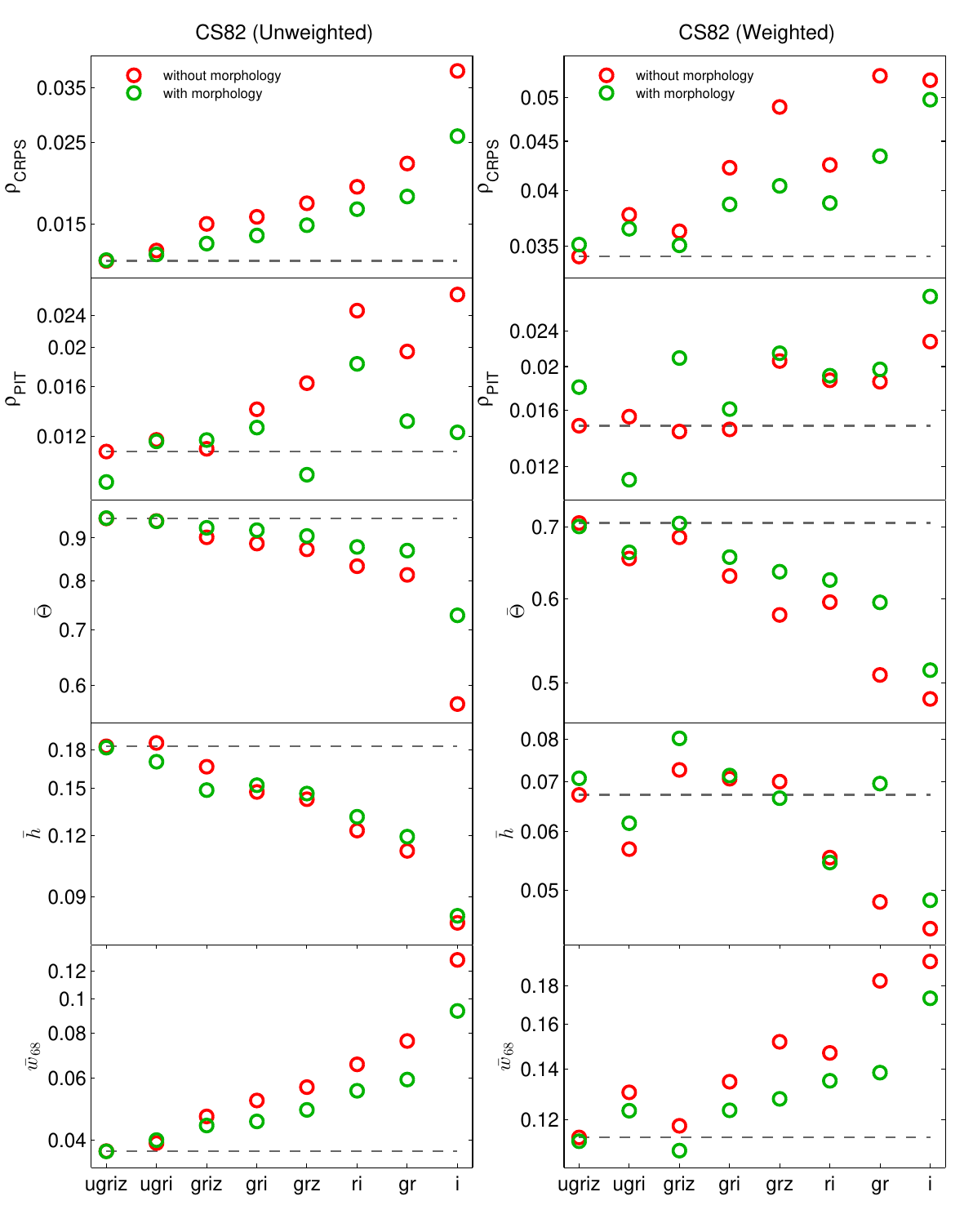}
\caption{Comparison of the mean CRPS ($\rho_\textrm{CRPS}$, first row) and root-mean-square PIT value ($\rho_\textrm{PIT}$, second row) when morphology is added to the training (green circles) on the unweighted (left) and weighted (right) CS82 training sample. Also shown are the mean odds ($\bar{\Theta}$, third row), mean peak height ($\bar{h}$, fourth row) and mean $68\%$ width ($\bar{w}_{68}$ last row) of the $p(z)$ produced for each case. The grey lines indicate the metric value of the pure $ugriz$ run.} \label{fig:res_pdfnz:pzmet}
\end{figure}

The values of $\rho_\textrm{CRPS}$, $\rho_\textrm{PIT}$, $\bar{\Theta}$, $\bar{h}$ and $\bar{w}_{68}$ are calculated for both the unweighted and weighted CS82 samples, for cases when trained with different numbers of magnitudes, with and without morphology, and the results are shown in Fig.~\ref{fig:res_pdfnz:pzmet}. Overall, we see that for both cases most metrics experience more improvement than degradation when morphology is added to training, and the trend of smaller improvement in with increasing number of bands used is also seen in most metrics. Notably $\rho_\textrm{CRPS}$, we see that galaxy morphology improves the forecast of the $p(z)$ immensely in both the weighted and unweighted samples, in most cases improvements of at least $5\%$ are achieved.

The change in $\rho_\textrm{PIT}$ is somewhat different from that of $\rho_\textrm{CRPS}$: while we see improvement in the unweighted sample, degradation in seen when the training sample is weighted. The reason for an increase in $\rho_\textrm{PIT}$ is not easy to pin down just by merely looking at the metric alone, since it could go both ways: either the widths of the $p(z)$ are going too narrow, or too wide. But with the values of $\bar{w}_{68}$ generally decreasing, we deduce that the sizes of the $p(z)$ have turned narrower than they should be with the inclusion of morphology.

In regards to $\bar{\Theta}$, for a $5$ $ugriz$ band training with morphology we find that the change in $\bar{\Theta}$ is almost negligible with morphology in the unweighted case: $\bar{\Theta}$ increased from $0.950$ to $0.951$ when multiple morphological parameters are included, partly because $\bar{\Theta}$ is very high to begin with. However we see a general improvement in both the unweighted and weighted case when morphology is added to the training, which indicates that morphology is indeed helping the ANN to improve the confidence in photo-$z$ values. We also see a direct correlation between the improvement of $\bar{\Theta}$ and $\rho_\textrm{CRPS}$, this is expected since both measure how well the expected redshift has been encapsulated within the PDF. However when checked for individual galaxies, we have verified that there is almost no correlation between the improvement in the point estimate photo-$z$ and the improvement in $\Theta$ for individual galaxies when morphology is added. This suggests that a galaxy with high $\Theta$ does not necessarily dictate a better photo-$z$ point estimate, although it remains sufficiently useful to remove outliers across an entire sample of galaxies \cite{soo_morpho-z:_2018}.

The change in $\bar{w}_{68}$ follows a similar trend as $\bar{\Theta}$, where we see that morphology on average has successfully reduced the widths of the $p(z)$. However, the impact of morphology on $\bar{h}$ shows a different story: there is a mixture of improvement and degradation. In fact, we find low correlation not only between $\bar{h}$ and the performance metrics across number of filters used, we also find low correlation between the improvement with $\bar{h}$ and the improvement in photo-$z$ for individual objects when morphology is added. We also do not see significant correlation between the improvement of $h$ and $\Theta$ either.

The motivation to study the change in $\bar{h}$ was in fact to see if morphology could decrease the degeneracy of multiple peaks, making only one peak stand out to obtain a more accurate photo-$z$. So to probe this a little further, we tabulated the distribution of $h$ for each run. From the distribution of PDF heights, we find that the shift in the peak of the $h$ distribution is in fact very small, and this is in contrast with the change in $w_{68}$ and $\Theta$. From further inspection of individual PDFs, we also find that many PDFs have very fuzzy and noisy distributions, some even having more than $10$ peaks in a single distribution. We suggest that these two factors have inhibited $\bar{h}$ from becoming a viable indicator for improvement in $p(z)$, and could be improved if smoother $p(z)$'s are produced.

\section{Conclusion and Future Work}

From our study, it is clear that $\rho_\textrm{PIT}$ is not a good metric to study improvement / degradation of the $p(z)$ in the context of CS82, although it remains an important metric to measure over- and under-confidences in PDFs. $\rho_\textrm{CRPS}$, $\bar{\Theta}$ and $\bar{w}_{68}$ are shown to be promising metrics to evaluate improvements in the \photoz PDFs reflected by the quality of point estimates. The viability of the metric $\bar{h}$ remains to be verified, and methods to produce smoother $p(z)$'s will be explored to achieve this purpose.

Note that in this work we have assumed that an improvement in the photo-$z$ estimates implies an improvement in the $p(z)$ and $n(z)$, and this assumption will be analysed in future work. Other future work include studies of $QQ$ plots, testing on improvements without relying on galaxy morphology, PDF smoothing methods and also the exploration of other metrics.

%%%%%%%%%%%%%%%%%%%%%%%
%%% ACKNOWLEDGEMENT %%%
%%%%%%%%%%%%%%%%%%%%%%%
\begin{acknowledgments}
JYHS acknowledges the financial support from the MyBrainSc Scholarship endowed by the Ministry of Education, Malaysia during which this work was completed. JYHS would also like to thank Ofer Lahav, Samuel Schmidt, Alex Malz and other members of the Large Synoptic Survey Telescope (LSST) Photo-$z$ Working Group for fruitful discussions which led to this work.
\end{acknowledgments}

%\nocite{*}
\bibliographystyle{apsrev4-2}
\bibliography{zaipsamp}% Produces the bibliography via BibTeX.

\providecommand{\noopsort}[1]{}\providecommand{\singleletter}[1]{#1}%
\begin{thebibliography}{14}%
\makeatletter
\providecommand \@ifxundefined [1]{%
 \@ifx{#1\undefined}
}%
\providecommand \@ifnum [1]{%
 \ifnum #1\expandafter \@firstoftwo
 \else \expandafter \@secondoftwo
 \fi
}%
\providecommand \@ifx [1]{%
 \ifx #1\expandafter \@firstoftwo
 \else \expandafter \@secondoftwo
 \fi
}%
\providecommand \natexlab [1]{#1}%
\providecommand \enquote  [1]{``#1''}%
\providecommand \bibnamefont  [1]{#1}%
\providecommand \bibfnamefont [1]{#1}%
\providecommand \citenamefont [1]{#1}%
\providecommand \href@noop [0]{\@secondoftwo}%
\providecommand \href [0]{\begingroup \@sanitize@url \@href}%
\providecommand \@href[1]{\@@startlink{#1}\@@href}%
\providecommand \@@href[1]{\endgroup#1\@@endlink}%
\providecommand \@sanitize@url [0]{\catcode `\\12\catcode `\$12\catcode
  `\&12\catcode `\#12\catcode `\^12\catcode `\_12\catcode `\%12\relax}%
\providecommand \@@startlink[1]{}%
\providecommand \@@endlink[0]{}%
\providecommand \url  [0]{\begingroup\@sanitize@url \@url }%
\providecommand \@url [1]{\endgroup\@href {#1}{\urlprefix }}%
\providecommand \urlprefix  [0]{URL }%
\providecommand \Eprint [0]{\href }%
\providecommand \doibase [0]{https://doi.org/}%
\providecommand \selectlanguage [0]{\@gobble}%
\providecommand \bibinfo  [0]{\@secondoftwo}%
\providecommand \bibfield  [0]{\@secondoftwo}%
\providecommand \translation [1]{[#1]}%
\providecommand \BibitemOpen [0]{}%
\providecommand \bibitemStop [0]{}%
\providecommand \bibitemNoStop [0]{.\EOS\space}%
\providecommand \EOS [0]{\spacefactor3000\relax}%
\providecommand \BibitemShut  [1]{\csname bibitem#1\endcsname}%
\let\auto@bib@innerbib\@empty
%</preamble>
\bibitem [{\citenamefont {Fern\'andez-Soto}\ \emph {et~al.}(2002)\citenamefont
  {Fern\'andez-Soto}, \citenamefont {Lanzetta}, \citenamefont {Chen},
  \citenamefont {Levine},\ and\ \citenamefont
  {Yahata}}]{fernandez-soto_error_2002}%
  \BibitemOpen
  \bibfield  {author} {\bibinfo {author} {\bibfnamefont {A.}~\bibnamefont
  {Fern\'andez-Soto}}, \bibinfo {author} {\bibfnamefont {K.~M.}\ \bibnamefont
  {Lanzetta}}, \bibinfo {author} {\bibfnamefont {H.-W.}\ \bibnamefont {Chen}},
  \bibinfo {author} {\bibfnamefont {B.}~\bibnamefont {Levine}},\ and\ \bibinfo
  {author} {\bibfnamefont {N.}~\bibnamefont {Yahata}},\ }\href
  {https://doi.org/10.1046/j.1365-8711.2002.05131.x} {\bibfield  {journal}
  {\bibinfo  {journal} {{MNRAS}}\ }\textbf {\bibinfo {volume} {330}},\ \bibinfo
  {pages} {889} (\bibinfo {year} {2002})}\BibitemShut {NoStop}%
\bibitem [{\citenamefont {Polsterer}\ \emph {et~al.}(2016)\citenamefont
  {Polsterer}, \citenamefont {D'Isanto},\ and\ \citenamefont
  {Gieseke}}]{polsterer_uncertain_2016}%
  \BibitemOpen
  \bibfield  {author} {\bibinfo {author} {\bibfnamefont {K.~L.}\ \bibnamefont
  {Polsterer}}, \bibinfo {author} {\bibfnamefont {A.}~\bibnamefont
  {D'Isanto}},\ and\ \bibinfo {author} {\bibfnamefont {F.}~\bibnamefont
  {Gieseke}},\ }\href@noop {} {\bibfield  {journal} {\bibinfo  {journal}
  {{arXiv} e-prints}\ } (\bibinfo {year} {2016})},\ \Eprint
  {https://arxiv.org/abs/1608.08016} {1608.08016} \BibitemShut {NoStop}%
\bibitem [{\citenamefont {Gerdes}\ \emph {et~al.}(2010)\citenamefont {Gerdes},
  \citenamefont {Sypniewski}, \citenamefont {{McKay}}, \citenamefont {Hao},
  \citenamefont {Weis}, \citenamefont {Wechsler},\ and\ \citenamefont
  {Busha}}]{gerdes_arborz:_2010}%
  \BibitemOpen
  \bibfield  {author} {\bibinfo {author} {\bibfnamefont {D.~W.}\ \bibnamefont
  {Gerdes}}, \bibinfo {author} {\bibfnamefont {A.~J.}\ \bibnamefont
  {Sypniewski}}, \bibinfo {author} {\bibfnamefont {T.~A.}\ \bibnamefont
  {{McKay}}}, \bibinfo {author} {\bibfnamefont {J.}~\bibnamefont {Hao}},
  \bibinfo {author} {\bibfnamefont {M.~R.}\ \bibnamefont {Weis}}, \bibinfo
  {author} {\bibfnamefont {R.~H.}\ \bibnamefont {Wechsler}},\ and\ \bibinfo
  {author} {\bibfnamefont {M.~T.}\ \bibnamefont {Busha}},\ }\href
  {https://doi.org/10.1088/0004-637X/715/2/823} {\bibfield  {journal} {\bibinfo
   {journal} {{ApJ}}\ }\textbf {\bibinfo {volume} {715}},\ \bibinfo {pages}
  {823} (\bibinfo {year} {2010})}\BibitemShut {NoStop}%
\bibitem [{\citenamefont {Bonnett}(2015)}]{bonnett_using_2015}%
  \BibitemOpen
  \bibfield  {author} {\bibinfo {author} {\bibfnamefont {C.}~\bibnamefont
  {Bonnett}},\ }\href {https://doi.org/10.1093/mnras/stv230} {\bibfield
  {journal} {\bibinfo  {journal} {{MNRAS}}\ }\textbf {\bibinfo {volume}
  {449}},\ \bibinfo {pages} {1043} (\bibinfo {year} {2015})}\BibitemShut
  {NoStop}%
\bibitem [{\citenamefont {Wittman}\ \emph {et~al.}(2016)\citenamefont
  {Wittman}, \citenamefont {Bhaskar},\ and\ \citenamefont
  {Tobin}}]{wittman_overconfidence_2016}%
  \BibitemOpen
  \bibfield  {author} {\bibinfo {author} {\bibfnamefont {D.~M.}\ \bibnamefont
  {Wittman}}, \bibinfo {author} {\bibfnamefont {R.}~\bibnamefont {Bhaskar}},\
  and\ \bibinfo {author} {\bibfnamefont {R.}~\bibnamefont {Tobin}},\ }\href
  {https://doi.org/10.1093/mnras/stw261} {\bibfield  {journal} {\bibinfo
  {journal} {{MNRAS}}\ }\textbf {\bibinfo {volume} {457}},\ \bibinfo {pages}
  {4005} (\bibinfo {year} {2016})}\BibitemShut {NoStop}%
\bibitem [{\citenamefont {Schmidt}\ \emph {et~al.}(prep)\citenamefont
  {Schmidt}, \citenamefont {Malz}, \citenamefont {Soo}, \citenamefont
  {Almosallam}, \citenamefont {Brescia}, \citenamefont {Cavuoti}, \citenamefont
  {Cohen-Tanugi}, \citenamefont {Connolly} \emph
  {et~al.}}]{schmidt_evaluation_nodate}%
  \BibitemOpen
  \bibfield  {author} {\bibinfo {author} {\bibfnamefont {S.~J.}\ \bibnamefont
  {Schmidt}}, \bibinfo {author} {\bibfnamefont {A.~I.}\ \bibnamefont {Malz}},
  \bibinfo {author} {\bibfnamefont {J.~Y.~H.}\ \bibnamefont {Soo}}, \bibinfo
  {author} {\bibfnamefont {I.~A.}\ \bibnamefont {Almosallam}}, \bibinfo
  {author} {\bibfnamefont {M.}~\bibnamefont {Brescia}}, \bibinfo {author}
  {\bibfnamefont {S.}~\bibnamefont {Cavuoti}}, \bibinfo {author} {\bibfnamefont
  {J.}~\bibnamefont {Cohen-Tanugi}}, \bibinfo {author} {\bibfnamefont {A.~J.}\
  \bibnamefont {Connolly}}, \emph {et~al.},\ }\href@noop {} {\bibfield
  {journal} {\bibinfo  {journal} {~}\ } (\bibinfo {year} {in
  prep.})}\BibitemShut {NoStop}%
\bibitem [{\citenamefont {Soo}\ \emph {et~al.}(2018)\citenamefont {Soo},
  \citenamefont {Moraes}, \citenamefont {Joachimi}, \citenamefont {Hartley},
  \citenamefont {Lahav}, \citenamefont {Charbonnier}, \citenamefont {Makler},
  \citenamefont {Pereira} \emph {et~al.}}]{soo_morpho-z:_2018}%
  \BibitemOpen
  \bibfield  {author} {\bibinfo {author} {\bibfnamefont {J.~Y.~H.}\
  \bibnamefont {Soo}}, \bibinfo {author} {\bibfnamefont {B.}~\bibnamefont
  {Moraes}}, \bibinfo {author} {\bibfnamefont {B.}~\bibnamefont {Joachimi}},
  \bibinfo {author} {\bibfnamefont {W.}~\bibnamefont {Hartley}}, \bibinfo
  {author} {\bibfnamefont {O.}~\bibnamefont {Lahav}}, \bibinfo {author}
  {\bibfnamefont {A.}~\bibnamefont {Charbonnier}}, \bibinfo {author}
  {\bibfnamefont {M.}~\bibnamefont {Makler}}, \bibinfo {author} {\bibfnamefont
  {M.~E.~S.}\ \bibnamefont {Pereira}}, \emph {et~al.},\ }\href
  {https://doi.org/10.1093/mnras/stx3201} {\bibfield  {journal} {\bibinfo
  {journal} {{MNRAS}}\ }\textbf {\bibinfo {volume} {475}},\ \bibinfo {pages}
  {3613} (\bibinfo {year} {2018})}\BibitemShut {NoStop}%
\bibitem [{\citenamefont {Annis}\ \emph {et~al.}(2014)\citenamefont {Annis},
  \citenamefont {Soares-Santos}, \citenamefont {Strauss}, \citenamefont
  {Becker}, \citenamefont {Dodelson}, \citenamefont {Fan}, \citenamefont
  {Gunn}, \citenamefont {Hao} \emph {et~al.}}]{annis_sloan_2014}%
  \BibitemOpen
  \bibfield  {author} {\bibinfo {author} {\bibfnamefont {J.}~\bibnamefont
  {Annis}}, \bibinfo {author} {\bibfnamefont {M.}~\bibnamefont
  {Soares-Santos}}, \bibinfo {author} {\bibfnamefont {M.~A.}\ \bibnamefont
  {Strauss}}, \bibinfo {author} {\bibfnamefont {A.~C.}\ \bibnamefont {Becker}},
  \bibinfo {author} {\bibfnamefont {S.}~\bibnamefont {Dodelson}}, \bibinfo
  {author} {\bibfnamefont {X.}~\bibnamefont {Fan}}, \bibinfo {author}
  {\bibfnamefont {J.~E.}\ \bibnamefont {Gunn}}, \bibinfo {author}
  {\bibfnamefont {J.}~\bibnamefont {Hao}}, \emph {et~al.},\ }\href
  {https://doi.org/10.1088/0004-637X/794/2/120} {\bibfield  {journal} {\bibinfo
   {journal} {{ApJ}}\ }\textbf {\bibinfo {volume} {794}},\ \bibinfo {pages}
  {120} (\bibinfo {year} {2014})}\BibitemShut {NoStop}%
\bibitem [{\citenamefont {Moraes}\ \emph {et~al.}(4 10)\citenamefont {Moraes},
  \citenamefont {Kneib}, \citenamefont {Leauthaud}, \citenamefont {Makler},
  \citenamefont {Van~Waerbeke}, \citenamefont {Bundy}, \citenamefont {Erben},
  \citenamefont {Heymans} \emph {et~al.}}]{moraes_cfht/megacam_2014}%
  \BibitemOpen
  \bibfield  {author} {\bibinfo {author} {\bibfnamefont {B.}~\bibnamefont
  {Moraes}}, \bibinfo {author} {\bibfnamefont {J.-P.}\ \bibnamefont {Kneib}},
  \bibinfo {author} {\bibfnamefont {A.}~\bibnamefont {Leauthaud}}, \bibinfo
  {author} {\bibfnamefont {M.}~\bibnamefont {Makler}}, \bibinfo {author}
  {\bibfnamefont {L.}~\bibnamefont {Van~Waerbeke}}, \bibinfo {author}
  {\bibfnamefont {K.}~\bibnamefont {Bundy}}, \bibinfo {author} {\bibfnamefont
  {T.}~\bibnamefont {Erben}}, \bibinfo {author} {\bibfnamefont
  {C.}~\bibnamefont {Heymans}}, \emph {et~al.},\ }in\ \href
  {http://www.astroscu.unam.mx/rmaa/RMxAC..44/PDF/RMxAC..44_ABSTRACTS.pdf}
  {\emph {\bibinfo {booktitle} {Revista Mexicana de Astronomia y Astrofisica
  Conference Series}}},\ \bibinfo {series} {Revista Mexicana de Astronomia y
  Astrofisica, vol. 27}, Vol.~\bibinfo {volume} {44}\ (\bibinfo {year}
  {2014-10})\ pp.\ \bibinfo {pages} {202--203}\BibitemShut {NoStop}%
\bibitem [{\citenamefont {York}\ \emph {et~al.}(2000)\citenamefont {York},
  \citenamefont {Adelman}, \citenamefont {Anderson}, \citenamefont {Anderson},
  \citenamefont {Annis}, \citenamefont {Bahcall}, \citenamefont {Bakken},
  \citenamefont {Barkhouser} \emph {et~al.}}]{york_sloan_2000}%
  \BibitemOpen
  \bibfield  {author} {\bibinfo {author} {\bibfnamefont {D.~G.}\ \bibnamefont
  {York}}, \bibinfo {author} {\bibfnamefont {J.}~\bibnamefont {Adelman}},
  \bibinfo {author} {\bibfnamefont {J.~E.~J.}\ \bibnamefont {Anderson}},
  \bibinfo {author} {\bibfnamefont {S.~F.}\ \bibnamefont {Anderson}}, \bibinfo
  {author} {\bibfnamefont {J.}~\bibnamefont {Annis}}, \bibinfo {author}
  {\bibfnamefont {N.~A.}\ \bibnamefont {Bahcall}}, \bibinfo {author}
  {\bibfnamefont {J.~A.}\ \bibnamefont {Bakken}}, \bibinfo {author}
  {\bibfnamefont {R.}~\bibnamefont {Barkhouser}}, \emph {et~al.},\ }\href
  {https://doi.org/10.1086/301513} {\bibfield  {journal} {\bibinfo  {journal}
  {{AJ}}\ }\textbf {\bibinfo {volume} {120}},\ \bibinfo {pages} {1579}
  (\bibinfo {year} {2000})}\BibitemShut {NoStop}%
\bibitem [{\citenamefont {Newman}\ \emph {et~al.}(2013)\citenamefont {Newman},
  \citenamefont {Cooper}, \citenamefont {Davis}, \citenamefont {Faber},
  \citenamefont {Coil}, \citenamefont {Guhathakurta}, \citenamefont {Koo},
  \citenamefont {Phillips} \emph {et~al.}}]{newman_deep2_2013}%
  \BibitemOpen
  \bibfield  {author} {\bibinfo {author} {\bibfnamefont {J.~A.}\ \bibnamefont
  {Newman}}, \bibinfo {author} {\bibfnamefont {M.~C.}\ \bibnamefont {Cooper}},
  \bibinfo {author} {\bibfnamefont {M.}~\bibnamefont {Davis}}, \bibinfo
  {author} {\bibfnamefont {S.~M.}\ \bibnamefont {Faber}}, \bibinfo {author}
  {\bibfnamefont {A.~L.}\ \bibnamefont {Coil}}, \bibinfo {author}
  {\bibfnamefont {P.}~\bibnamefont {Guhathakurta}}, \bibinfo {author}
  {\bibfnamefont {D.~C.}\ \bibnamefont {Koo}}, \bibinfo {author} {\bibfnamefont
  {A.}~\bibnamefont {Phillips}}, \emph {et~al.},\ }\href
  {https://doi.org/10.1088/0067-0049/208/1/5} {\bibfield  {journal} {\bibinfo
  {journal} {{ApJS}}\ }\textbf {\bibinfo {volume} {208}},\ \bibinfo {pages} {5}
  (\bibinfo {year} {2013})}\BibitemShut {NoStop}%
\bibitem [{\citenamefont {Drinkwater}\ \emph {et~al.}(2010)\citenamefont
  {Drinkwater}, \citenamefont {Jurek}, \citenamefont {Blake}, \citenamefont
  {Woods}, \citenamefont {Pimbblet}, \citenamefont {Glazebrook}, \citenamefont
  {Sharp}, \citenamefont {Pracy} \emph {et~al.}}]{drinkwater_wigglez_2010}%
  \BibitemOpen
  \bibfield  {author} {\bibinfo {author} {\bibfnamefont {M.~J.}\ \bibnamefont
  {Drinkwater}}, \bibinfo {author} {\bibfnamefont {R.~J.}\ \bibnamefont
  {Jurek}}, \bibinfo {author} {\bibfnamefont {C.}~\bibnamefont {Blake}},
  \bibinfo {author} {\bibfnamefont {D.}~\bibnamefont {Woods}}, \bibinfo
  {author} {\bibfnamefont {K.}~\bibnamefont {Pimbblet}}, \bibinfo {author}
  {\bibfnamefont {K.}~\bibnamefont {Glazebrook}}, \bibinfo {author}
  {\bibfnamefont {R.}~\bibnamefont {Sharp}}, \bibinfo {author} {\bibfnamefont
  {M.~B.}\ \bibnamefont {Pracy}}, \emph {et~al.},\ }\href
  {https://doi.org/10.1111/j.1365-2966.2009.15754.x} {\bibfield  {journal}
  {\bibinfo  {journal} {{MNRAS}}\ }\textbf {\bibinfo {volume} {401}},\ \bibinfo
  {pages} {1429} (\bibinfo {year} {2010})}\BibitemShut {NoStop}%
\bibitem [{\citenamefont {Le~F\`evre}\ \emph {et~al.}(2013)\citenamefont
  {Le~F\`evre}, \citenamefont {Cassata}, \citenamefont {Cucciati},
  \citenamefont {Garilli}, \citenamefont {Ilbert}, \citenamefont {Le~Brun},
  \citenamefont {Maccagni}, \citenamefont {Moreau} \emph
  {et~al.}}]{le_fevre_vimos_2013}%
  \BibitemOpen
  \bibfield  {author} {\bibinfo {author} {\bibfnamefont {O.}~\bibnamefont
  {Le~F\`evre}}, \bibinfo {author} {\bibfnamefont {P.}~\bibnamefont {Cassata}},
  \bibinfo {author} {\bibfnamefont {O.}~\bibnamefont {Cucciati}}, \bibinfo
  {author} {\bibfnamefont {B.}~\bibnamefont {Garilli}}, \bibinfo {author}
  {\bibfnamefont {O.}~\bibnamefont {Ilbert}}, \bibinfo {author} {\bibfnamefont
  {V.}~\bibnamefont {Le~Brun}}, \bibinfo {author} {\bibfnamefont
  {D.}~\bibnamefont {Maccagni}}, \bibinfo {author} {\bibfnamefont
  {C.}~\bibnamefont {Moreau}}, \emph {et~al.},\ }\href
  {https://doi.org/10.1051/0004-6361/201322179} {\bibfield  {journal} {\bibinfo
   {journal} {A\&A}\ }\textbf {\bibinfo {volume} {559}},\ \bibinfo {pages} {21}
  (\bibinfo {year} {2013})}\BibitemShut {NoStop}%
\bibitem [{\citenamefont {Sadeh}\ \emph {et~al.}(2016)\citenamefont {Sadeh},
  \citenamefont {Abdalla},\ and\ \citenamefont {Lahav}}]{sadeh_annz2:_2016}%
  \BibitemOpen
  \bibfield  {author} {\bibinfo {author} {\bibfnamefont {I.}~\bibnamefont
  {Sadeh}}, \bibinfo {author} {\bibfnamefont {F.~B.}\ \bibnamefont {Abdalla}},\
  and\ \bibinfo {author} {\bibfnamefont {O.}~\bibnamefont {Lahav}},\ }\href
  {https://doi.org/10.1088/1538-3873/128/968/104502} {\bibfield  {journal}
  {\bibinfo  {journal} {{PASP}}\ }\textbf {\bibinfo {volume} {128}},\ \bibinfo
  {pages} {104502} (\bibinfo {year} {2016})}\BibitemShut {NoStop}%
\end{thebibliography}%

\end{document}